\documentclass[aps,prB,reprint,superscriptaddress,onecolumn]{revtex4-1}
\usepackage{amsmath,amssymb}
\usepackage{graphicx}
\usepackage{color}
\draft

\begin{document}
\makeatletter 
\makeatother

\title{Coulomb collisions of hot and cold single electrons in series-coupled silicon single-electron pumps}
\author{Gento Yamahata}
\email{E-mail: gento.yamahata@ntt.com}
\affiliation{NTT Basic Research Laboratories, NTT Corporation, 3-1 Morinosato Wakamiya, Atsugi, Kanagawa 243-0198, Japan}

\author{Nathan Johnson\footnote{Present address: London Centre for Nanotechnology, University College London, London WC1H 0AH, United Kingdom}}
\affiliation{NTT Basic Research Laboratories, NTT Corporation, 3-1 Morinosato Wakamiya, Atsugi, Kanagawa 243-0198, Japan}

\author{Akira Fujiwara}
\affiliation{NTT Basic Research Laboratories, NTT Corporation, 3-1 Morinosato Wakamiya, Atsugi, Kanagawa 243-0198, Japan}
\begin{abstract}
Control of the Coulomb interaction between single electrons is vital for realizing quantum information processing using flying electrons and, particularly, for the realization of deterministic two-qubit operations. Since the strength of the Coulomb interaction increases with decreasing distance, a collision experiment of single electrons would be an ideal way to investigate it. Moreover, it would be useful to study such a Coulomb collision in silicon system, which has been extensively studied for qubit applications but so far has not been used for making Coulomb collisions at the single-electron level. Here, we made two series-coupled tunable-barrier single-electron pumps in silicon and used one to inject a hot single electron into the other pump in each pumping cycle. The hot single electron collides with a cold single electron confined in the other single-electron pump. We observed a current flow due to ejection not only of the hot single electron but also of the confined cold single electron. The latter leads to an excess current at a current plateau at a certain voltage range. We also found that increasing the number of cold electrons from one to two increased the cold-electron current by at least twofold. These results can be explained by a charging effect due to the Coulomb interaction. This observation is an important step toward quantum manipulation of flying single electrons in silicon.       
\end{abstract}
\maketitle
\section{Introduction}
A single-electron (SE) pump using a clock-controlled dynamic quantum dot (QD) can accurately emit hot electrons one by one \cite{tunable-barrier1}. It can be used for applications to quantum current standards \cite{pekola-rev, UnivRev}, quantum information technology \cite{SEsouce1, GYnnano}, quantum sensing \cite{Nathan_samp}, and electron quantum optics \cite{Ubbelohde1,Jtom}. In particular, the emitted hot SE propagating in a solid-state device is one of the key candidates for quantum information processing using flying SEs \cite{FlyRev}. This type of flying qubit has recently attracted attention \cite{yamamoto_fly, takadaSAW1,ihoFly,shimizu50} as a counterpart of photonic quantum computing \cite{photo1, photo2} because, as opposed to photons, SEs inherently have the Coulomb interaction, with which a deterministic two-qubit gate is expected to be realized \cite{Fly2qb}. However, a two-electron collision experiment on SEs propagating in the quantum Hall edge channel at energies close to the Fermi level has shown that the Coulomb interaction is negligible due to screening by the many surrounding electrons \cite{Feve_sci_coh}. On the other hand, more recently, theoretical works \cite{HOMci,Sbarrier,EliColl} have pointed out an importance of the Coulomb interaction. Furthermore, the unscreened Coulomb interaction has been experimentally observe between two flying SEs generated by GaAs SE pumps \cite{Jcol,Ncol} and by surface acoustic waves in a GaAs device \cite{SAWcol}. These results would indicate that the unscreened devices are important for achieving a controllable flying-qubit gate.

An attractive candidate for the SE source for the flying qubit is a silicon SE pump, as it is capable of high-accuracy and high-speed operation \cite{NPL-NTT1,Zhao_pump} even at liquid helium temperatures \cite{LHeSi}. Such relatively high-temperature operation is desirable for a flying qubit initialization because a cooling power in the millikelvin regime is limited, which is a general problem of most solid-state qubit systems \cite{1KSiqubit,1KsiDel}. In addition, the fact that silicon has weak spin-orbit and hyperfine interactions would be an advantage for stable propagation of hot SEs. Furthermore, silicon QDs have been extensively studied toward static-qubit applications, thanks to the long coherence time of electron spins confined in the QDs and the widespread availability of silicon integration technology \cite{NoiriUniv,SurDel,TakedaEC,Si6qubit}. Combining them with flying electrons would be another important pathway toward realization of quantum information devices \cite{MeunSAW1,MeunSAW2}.

Considering the above background, it is valuable to investigate the Coulomb interaction of flying SEs in silicon. So far, there have been investigations on the Coulomb interaction between a hot electron and many electrons in the Fermi sea of silicon devices \cite{NEChot, OnoHot} and amplification of current due to the Coulomb interaction has been reported \cite{OnoHot}. However, there is no report on the Coulomb interaction of flying SEs at the SE level. Here, we propose to use two SE pumps based on silicon QDs that are connected in series for collision experiments between a hot flying SE and a cold SE trapped in one of the QDs. We observed an excess current at the current plateau of the double SE pumps that was evidence of the unscreened Coulomb interaction.

The outline of this paper is as follows. We introduce the device structure, its fabrication process, and the measurement scheme of the SE-collision experiments in Sec.~II. Then, we describe and discuss the experimental results in Sec.~III. In particular, Sec.~III-A discusses the current transport characteristics of the individual SE pumps and shows that each pump operated accurately. Section III-B describes the injection of hot SEs into a zero-electron QD and shows that the injected hot SE can be ejected over the QD in every cycle. Section III-C describes hot-SE injection into a one-electron QD, and an experimentally observed increase in current originating from the cold SE trapped in the QD due to the Coulomb interaction. Section III-D explores hot-SE injection into a two-electron QD, for which the experimental results were consistent with the one-electron case. Section III-E is an additional discussion. Section IV is the conclusion.

\section{Device and measurement scheme}
\begin{figure}
\begin{center}
\includegraphics[pagebox=artbox]{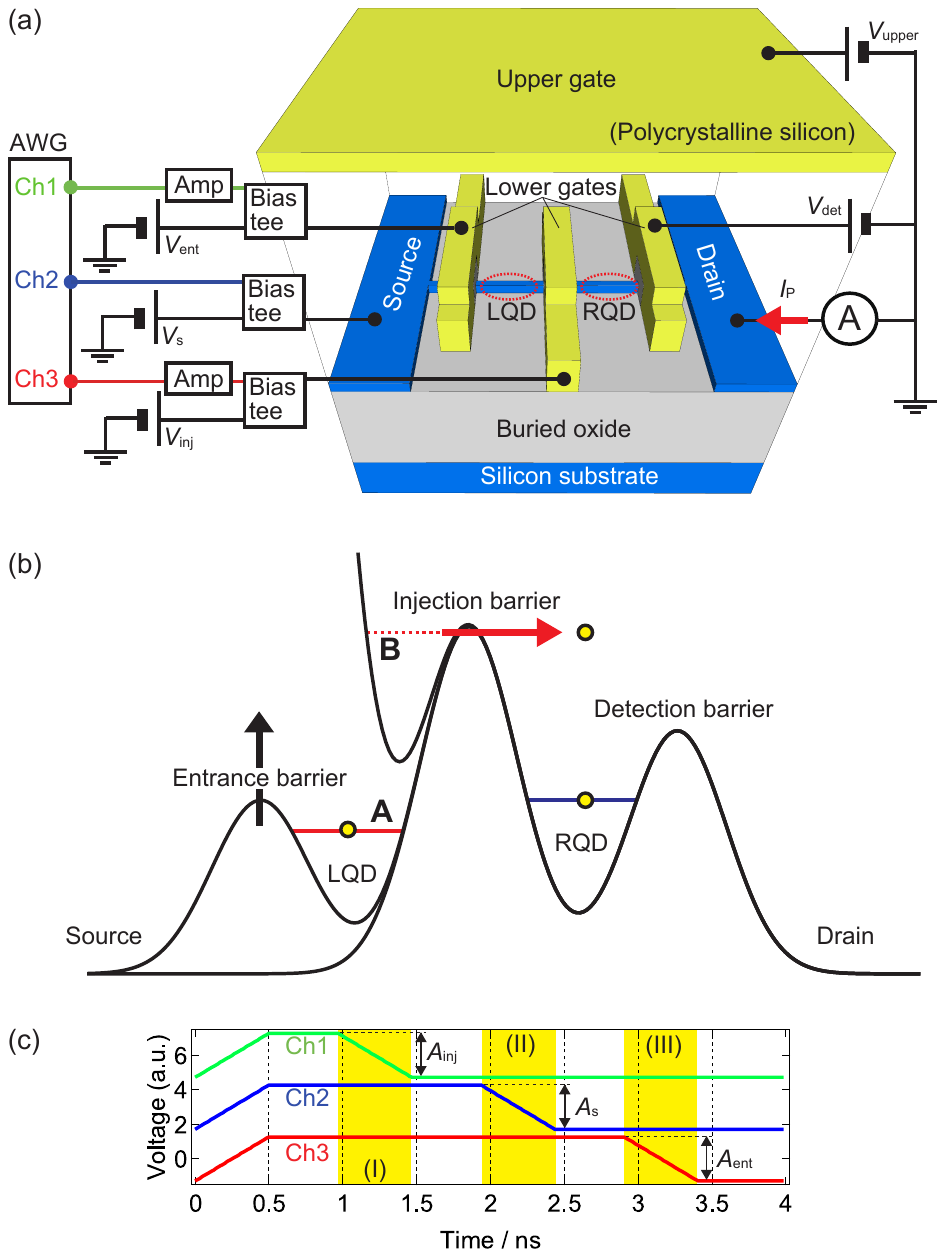}%width=230pt 
\end{center}
\caption{(a) Schematic device structure. The left (right) quantum dot [LQD (RQD)] is formed between the left (right) and center lower gates. The upper gate covers the whole region of the silicon wire and a part of the source and drain regions. DC voltages are applied to the upper gate ($V_{\mathrm{upper}}$) and the right lower gate ($V_{\mathrm{det}}$). High-frequency signals generated by an arbitrary waveform generator (AWG; Keysight M8195A) combined with DC voltages using bias tees are applied to the center lower gate (channel 1 of the AWG and $V_{\mathrm{inj}}$), source electrode (channel 2 of the AWG and $V_{\mathrm{s}}$), and left lower gate (channel 3 of the AWG and $V_{\mathrm{ent}}$). The outputs of channels 1 and 3 are amplified by a 15-GHz low-noise amplifier (Tektronix PSPL8003) with a nominal gain of 15 dB. A $-2$ dB attenuator is connected after the bias tee for channel 1 and $-3$ dB attenuators are set before the bias tee for channels 2 and 3 (not shown). All DC voltages are supplied by DC voltage sources (Yokogawa GS200). The DC current $I_{\mathrm{P}}$ at the drain terminal is converted into a DC voltage using a programmable current amplifier (NF CA5351) and the output DC voltage is measured using a digital multimeter (Keysight 3458A). (b) Schematic potential diagram during a collision measurement. The yellow dots are electrons. The number of electrons in the RQD is initialized to one in this case. An SE from the source electrode is captured by the LQD (condition A) and eventually injected as a hot SE into the RQD (condition B) with raising entrance barrier. (c) Pulse sequence for the collision measurement. $A_{\mathrm{inj}}$, $A_{\mathrm{s}}$, and $A_{\mathrm{ent}}$ are the amplitudes of the signal for channel 1, 2, and 3, respectively (output impedance is 50 ohms). The frequency $f$ is $250$ MHz. All rise and fall times are 0.5 ns.}
\label{f1}
 \end{figure}

Figure~\ref{f1}(a) shows the schematic device structure together with the electrical connections \cite{frat1,NPL-NTT1,GYnnano}. The fabrication process is as follows. A non-doped silicon wire is formed on 400-nm-thick buried oxide by using electron beam lithography and dry etching, followed by thermal oxidation for forming 30-nm-thick silicon dioxide. Next, n-type polycrystalline silicon is grown by chemical vapor deposition. This layer is patterned using electron beam lithography and dry etching to form the three lower gates of the device on the silicon wire. After growth of inter-layer silicon dioxide to a thickness of 50 nm by chemical vapor deposition, n-type polycrystalline silicon is again grown by chemical vapor deposition. This layer is patterned using optical lithography and dry etching to form an upper gate covering the whole region of the silicon wire.  Then, the upper gate is used as a mask for ion implantation to form n-type source and drain electrodes. Finally, aluminum ohmic contacts to the source, drain, and gate electrodes are formed using vacuum deposition. The diameter of the silicon wire, lower gate lengths, and spacing between adjacent lower gates are 15, 10, and 100 nm, respectively.

The device was cooled in a dilution refrigerator at a base temperature of 20 mK unless otherwise noted. Application of a positive DC voltage to the upper gate ($V_{\mathrm{upper}}=1.2$ V) induced electrons in the source and drain electrodes. We formed entrance, injection, and detection barriers in the silicon wire under the lower gates [see the potential diagram in Fig.~\ref{f1}(b)] by applying DC voltages to the left ($V_{\mathrm{ent}}$), center ($V_{\mathrm{inj}}$), and right ($V_{\mathrm{det}}$) lower gates, respectively. This resulted in the formation of the left and right QDs, which will be referred to as the LQD and RQD, respectively. 

The voltage pulse scheme shown in Fig.~\ref{f1}(c) enables the SE collision experiments for investigation of the Coulomb interaction between the hot SE injected from the LQD and the cold SE confined in the RQD. Here, three synchronized high-frequency signals generated by an arbitrary waveform generator are applied to the center lower gate, source, and left lower gate, with amplitudes of $A_{\mathrm{inj}}$, $A_{\mathrm{s}}$, and $A_{\mathrm{ent}}$, respectively. First, the entrance and injection barriers are lowered and the number of SEs in the RQD is set by raising the injection barrier [period I in Fig.~\ref{f1}(c)]. In this process, some electrons that are loaded from the source into the RQD (loading stage) escape back to the source during the rise of the barrier and then $n$ SEs are captured by the RQD (capture stage), followed by ejection of the captured SEs to the drain (ejection stage) \cite{kaestner1, tunable-barrier1, GYgval, GYnnano}. This tunable-barrier SE pumping using the RQD is used to reset the number of SEs in the RQD in each cycle. The reset in each cycle is necessary because subsequent injection of an SE from the LQD might increase the number of SEs in the RQD. The source voltage is used to control the relative number of SEs captured by the LQD and RQD [period II in Fig.~\ref{f1}(c)]. Finally, a hot SE is injected from the LQD to the RQD by raising the entrance barrier [period III in Fig.~\ref{f1}(c)]. This process is also tunable-barrier SE pumping using the LQD. 

The DC current ($I_{\mathrm{P}}$) flowing through the silicon wire was measured at the drain terminal during the above double SE pumping operation. The frequency $f$ of the signals was 250 MHz for all SE pumping measurements. For $n$ electrons transferred in a cycle, $I_{\mathrm{P}}=nef \sim 40\times n$ pA, where $e$ is the elementary charge. We applied a DC voltage to the source electrode ($V_{\mathrm{s}}$) only for the DC transport measurements ($V_{\mathrm{s}}=0$ V for the SE pumping). 

\section{Experimental results and discussions}
\subsection{Individual SE pump characteristics}
First, the individual SE pump characteristics were checked by using only the LQD or only the RQD. To transfer SEs using the LQD (RQD), we applied a positive DC voltage to the right (left) gate to sufficiently lower the detection (entrance) barrier. Then, we applied a high-frequency signal only to the left (center) gate for the LQD (RQD) pump. The pulse shape was identical to that shown in Fig.~\ref{f1}(c). 

\begin{figure}
\begin{center}
\includegraphics[pagebox=artbox]{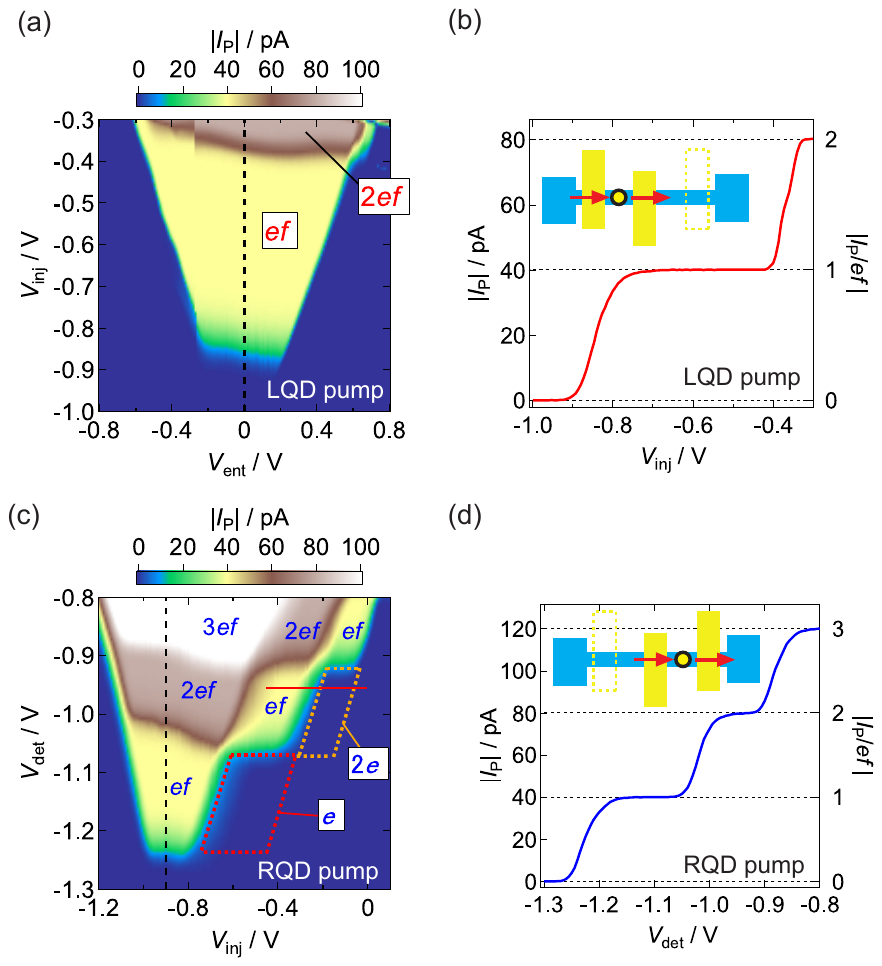}%width=230pt 
\end{center}
\caption{(a) Absolute value of current $I_{\mathrm{P}}$ as a function of $V_{\mathrm{ent}}$ and $V_{\mathrm{inj}}$ at $f=250$ MHz, where $V_{\mathrm{upper}}=1.2$ V, $V_{\mathrm{det}}=1$ V,  and $V_{\mathrm{s}}=0$ V. The high-frequency signal with an amplitude $A_{\mathrm{ent}}=0.3$ V is applied to the left gate. (b) Absolute value of current $I_{\mathrm{P}}$ (left axis) and $|I_{\mathrm{P}}/ef|$ (right axis) as a function of $V_{\mathrm{inj}}$ at $f=250$ MHz, where $V_{\mathrm{upper}}=1.2$ V, $V_{\mathrm{det}}=1$ V, $V_{\mathrm{ent}}=0$ V, $V_{\mathrm{s}}=0$ V, and $A_{\mathrm{ent}}=0.3$ V, corresponding to the black dashed line in (a). The inset is a schematic diagram of the LQD pump. (c) Absolute value of current $I_{\mathrm{P}}$ as a function of $V_{\mathrm{inj}}$ and $V_{\mathrm{det}}$ at $f=250$ MHz, where $V_{\mathrm{upper}}=1.2$ V, $V_{\mathrm{ent}}=1$ V,  and $V_{\mathrm{s}}=0$ V. The high-frequency signal with an amplitude $A_{\mathrm{inj}}=0.3$ V is applied to the center gate. The areas within the red and orange dotted lines are the values of $V_{\mathrm{inj}}$ and $V_{\mathrm{det}}$ at which one and two SEs remain in the RQD, respectively, when the height of the injection barrier is at its maximum. (d) Absolute value of current $I_{\mathrm{P}}$ (left axis) and $|I_{\mathrm{P}}/ef|$ (right axis) as a function of $V_{\mathrm{det}}$ at $f=250$ MHz, where $V_{\mathrm{upper}}=1.2$ V, $V_{\mathrm{ent}}=1$ V, $V_{\mathrm{inj}}=-0.9$ V, $V_{\mathrm{s}}=0$ V, and $A_{\mathrm{inj}}=0.3$ V, corresponding to the black dashed line in (c). The inset is a schematic diagram of the RQD pump.}
\label{f2}
 \end{figure}

Figure~\ref{f2}(a) shows $|I_{\mathrm{P}}|$ as a function of $V_{\mathrm{ent}}$ and $V_{\mathrm{inj}}$ when only the LQD pump was operated. The left, bottom, and right boundaries of the trapezoidal current-generating region correspond to loading, capture, and ejection of SEs, respectively. The current along the black dashed line on the LQD pump map at $V_{\mathrm{ent}}=0$ in Fig.~\ref{f2}(a) has a clear current plateau with a level of $ef$ [Fig.~\ref{f2}(b)]. This indicates that a hot SE can be accurately injected from the LQD.

Figure~\ref{f2}(c) shows $|I_{\mathrm{P}}|$ as a function of $V_{\mathrm{inj}}$ and $V_{\mathrm{det}}$ when only the RQD pump was operated. The current along the black dashed line on the RQD pump at $V_{\mathrm{inj}}=-0.9$ V in Fig.~\ref{f2}(c) has clear plateaus with levels of $ef$ and $2ef$ [Fig.~\ref{f2}(d)]. This indicates that the RQD can accurately capture one or two SEs. For the collision experiments, it is necessary that SEs remain captured by the RQD. The red or orange dotted squares in Fig.~\ref{f2}(c) are regions of $V_{\mathrm{inj}}$ and $V_{\mathrm{det}}$ where one or two SEs can be captured with high accuracy in the capture stage but the detection barrier prevents them from being ejected to the drain after the ejection stage. We focused on these regions in the following measurements.

\subsection{Hot-SE injection into a zero-electron RQD}
\begin{figure}
\begin{center}
\includegraphics[pagebox=artbox]{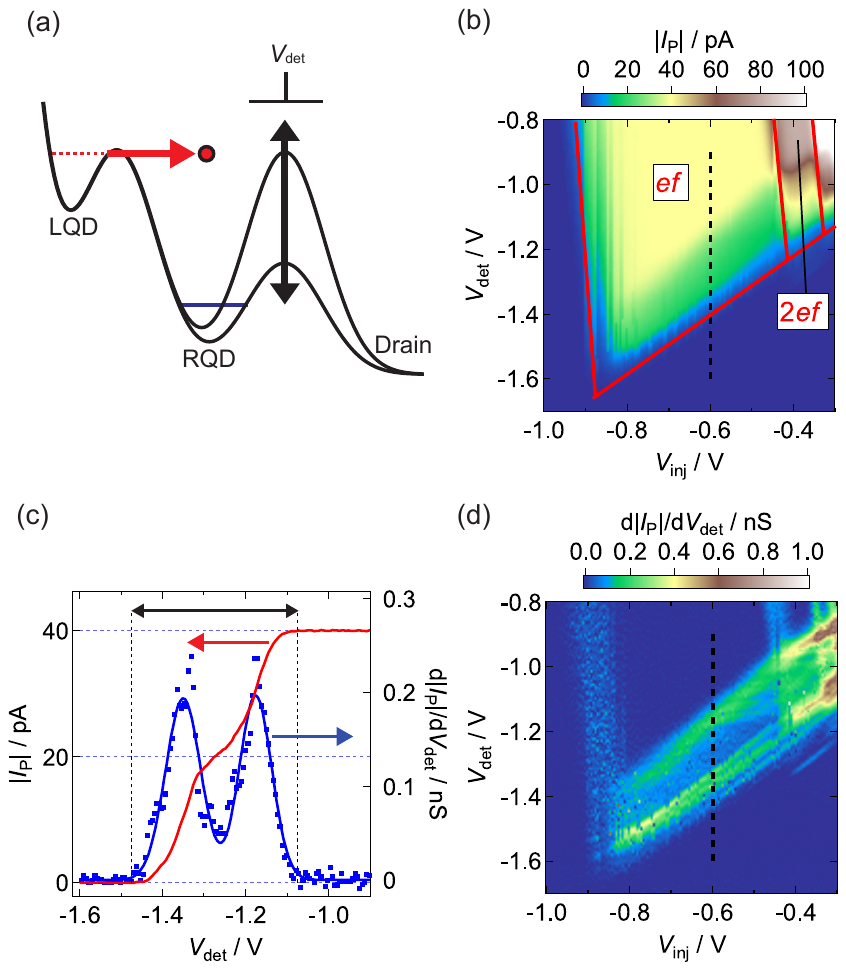}%width=230pt 
\end{center}
\caption{(a) Schematic potential diagram during a hot-SE injection from the LQD into zero-electron RQD. When $V_{\mathrm{det}}$ changes, the height of the detection barrier changes. This change is used for detection of the hot SE. (b) $|I_{\mathrm{P}}|$ as a function of $V_{\mathrm{inj}}$ and $V_{\mathrm{det}}$ with the pulse sequence shown in Fig.~\ref{f1}(c), where $V_{\mathrm{upper}}=1.2$ V, $V_{\mathrm{ent}}=-0.5$ V, $V_{\mathrm{s}}=0$ V, $A_{\mathrm{ent}}=0.3$ V, $A_{\mathrm{inj}}=0.075$ V, and $A_{\mathrm{s}}=0.5$ V. (c) $|I_{\mathrm{P}}|$ (red line; left axis) and $\mathrm{d}|I_{\mathrm{P}}|/\mathrm{d}V_{\mathrm{det}}$ (blue dots; right axis) as a function of $V_{\mathrm{det}}$, which corresponds to the black dashed lines in (b) and (d), respectively, where $V_{\mathrm{ent}}=-0.6$ V. The fit (blue line) to the blue dots is composed of two Gaussian functions: $A\mathrm{exp}[-\{(V_{\mathrm{det}}-V_{1})/w_{1}\}^2]+B\mathrm{exp}[-\{(V_{\mathrm{det}}-V_{2})/w_{2}\}^2]$, where $V_{1}$, $V_{2}$, $w_{1}$, $w_{2}$, $A$, and $B$ are constants. $\Delta V_{\mathrm{det}}=V_{2}-V_{1}\sim 0.17$ V. (d) Absolute value of the first derivative of $|I_{\mathrm{P}}|$ shown in (b).}
\label{f3}
 \end{figure}

Next, we investigated injection of a hot SE from the LQD into a zero-electron RQD [Fig.~\ref{f3}(a)] by using the pulse scheme shown in Fig.~\ref{f1}(c). Note that we define the situation in which no electrons are captured in the RQD during the loading or capture stages as ``zero-electron RQD''. In this case, $A_{\mathrm{inj}}$ is minimized to suppress the loading of electrons to the RQD. When the energy of the injected hot SE is lower than the height of the detection barrier, which is tuned by $V_{\mathrm{det}}$, the hot SE can not be ejected to the drain. This suppresses $I_{\mathrm{P}}$, and the energy of the injected hot SE with respect to the detection barrier top can be estimated. Figure~\ref{f3}(b) shows $|I_{\mathrm{P}}|$ as a function of $V_{\mathrm{inj}}$ and $V_{\mathrm{det}}$ in the case of the hot-SE injection to the zero-electron RQD. The three red lines falling to the right indicate threshold voltages for the capture stage of the LQD during the rise of the entrance barrier. From left to right in Fig.~\ref{f3}(b), the number of SEs captured by the LQD increases similar to the black dashed line in Fig.~\ref{f2}(a). The slight tilt of these red lines indicates a small capacitive coupling between the right lower gate and the LQD. The red line rising to the right indicates the threshold voltage for hot-SE ejection to the drain. Note that we used voltage conditions in which the ejection and loading lines of the LQD do not appear in the figure.

The red line in Fig.~\ref{f3}(c) corresponds to the blacked dashed line on the pump map in Fig.~\ref{f3}(b). The current plateau is at the level of $ef \sim 40$ pA at around $V_{\mathrm{det}}=-1$ V. This indicates that the hot SE injected from the LQD was ejected to the drain in every cycle. In addition, there is a clear two-step feature in $|I_{\mathrm{P}}|$. To examine it, we differentiated $|I_{\mathrm{P}}|$ with respect to $V_{\mathrm{det}}$ [blue dots in Fig.~\ref{f3}(c)]. A fit to the $\mathrm{d}|I_{\mathrm{P}}|/\mathrm{d}V_{\mathrm{det}}$ data in Fig.~\ref{f3}(c) using two Gaussian functions roughly reveals the spacing of these two peaks, $\Delta V_{\mathrm{det}} \sim 0.17$ V.

To convert $V_{\mathrm{det}}$ into a hot-SE energy relative to the top of the detection barrier, we estimated the voltage-to-energy conversion factors between the right lower gate and the detection barrier ($\alpha _{\mathrm{detB}} = 0.40$ eV/V) and between the right lower gate and the injection barrier ($\alpha _{\mathrm{det\_injB}} \sim 0.034$ eV/V) from DC measurements (Appendix A). The latter estimate is necessary because the energy of the injected hot SE is determined by the height of the injection barrier. Here, the minimum energy scale required for the hot-SE ejection to the drain in every cycle roughly corresponds to the length of the black arrow in Fig.~\ref{f3}(c), which is about 0.4 V. This value is converted into an energy of 0.4 V $\times (\alpha _{\mathrm{detB}}-\alpha _{\mathrm{det\_injB}}) \sim 0.15$ eV and is considered to be a typical hot-SE energy in this experiment. Then, the spacing $\Delta V_{\mathrm{det}}$ was converted into an energy: $\Delta V_{\mathrm{det}} \times (\alpha _{\mathrm{detB}}-\alpha _{\mathrm{det\_injB}}) \sim 62$ meV.

The lines of peaks on the pump map are parallel with an energy difference of about 62 meV [Fig.~\ref{f3}(d)]. Such parallel peak structures have been reported in hot-electron injection experiments on GaAs devices with quantum point contacts \cite{fujiHot} and SE pumps \cite{Jon-prl1,LarsCount}, in which the spacing is understood to be a result of relaxation of an electron due to LO-phonon emission with an energy of 36 meV. On the other hand, previous hot-electron injection experiments in silicon devices have not revealed evidence of phonon relaxation \cite{NEChot, OnoHot} probably because relaxation due to the Coulomb interaction between a hot electron and many electrons in the Fermi sea is much faster than phonon relaxation. In the case of silicon, which is a non-polar semiconductor with multiple conduction band valleys, inter-valley phonon scattering is important \cite{hamaguchi}. The largest contribution would be from what is called a g-LO phonon with an energy of 61-63 meV \cite{DiffG,takagiG,hamaguchi}, which is consistent with our calculated spacing. Although the two observed peaks might originate from the g-LO phonon emission, we need to conduct further experiments on several different devices for checking reproducibility of these lines before we can conclude what the parallel lines are attributed to.

\subsection{Hot-SE injection into a one-electron RQD}
Next, we examined injection of a hot SE from the LQD into the RQD with one electron in it. Figure~\ref{f4}(a) shows $|I_{\mathrm{P}}|$ for such a case as a function of $V_{\mathrm{inj}}$ and $V_{\mathrm{det}}$. The $ef$ and $2ef$ plateaus in the left side on the map correspond to the RQD-pump current. At the voltage conditions inside the red dashed square, there is one electron in the RQD after the injection barrier height is raised [period I in Fig.~\ref{f1}(c)]. In addition, current flows due to the injected hot SE from the LQD at the voltage conditions inside red lines. In the area within both the red lines and the red dashed square, the injected hot SE can collide with the SE confined in the RQD.

To investigate this area in more detail, Figs.~\ref{f4}(b) and (c) plot $|I_{\mathrm{P}}|$ along the horizontal and tilted black dashed lines in Fig.~\ref{f4}(a). The tilted voltage line is called $V_{\mathrm{tilt}}$, which is parallel to the threshold line determined by ejection of the SE from the RQD to the drain  [the left red dashed line in Fig.~\ref{f4}(a)]. This means that the height of the detection barrier with respect to the RQD is almost constant along $V_{\mathrm{tilt}}$ and the ejection probability of the SE captured by the RQD is almost constant. Note that both $V_{\mathrm{inj}}$ and $V_{\mathrm{det}}$ change along $V_{\mathrm{tilt}}$ but we have used $V_{\mathrm{inj}}$ for the horizontal axis in Fig~\ref{f4}(c). The intersection point of these two lines corresponds to the vertical dashed lines in Figs.~\ref{f4}(b) and (c). Around the intersection point, there is an excess current $\Delta I_{\mathrm{P}}\sim 2.2$ pA. The fact that $|I_{\mathrm{P}}|$ saturates in both plots indicates that the hot SE injected from the LQD was ejected to the drain in every cycle. Thus, $\Delta I_{\mathrm{P}}$ originates from the SE confined in the RQD.

Now let us explain how the SE confined in the RQD is ejected and contributes to $\Delta I_{\mathrm{P}}$ in more detail by using the potential diagrams shown in Figs.~\ref{f4}(d)-(g). As a first approximation, we consider a single-particle picture with a constant charging energy. Just before the hot SE is injected into the RQD, the LQD and RQD each have one electron [Fig.~\ref{f4}(d)]. Both electrons are assumed to occupy the ground state ($E_{\mathrm{G(L1)}}$ and $E_{\mathrm{G(R1)}}$). Note that there is a small possibility of nonadiabatic excitation, but its energy scale should be much smaller than the charging energy and hot-SE energy \cite{GYnnano}.

Then, the SE is injected from the LQD to RQD [Fig.~\ref{f4}(e)]. At that moment, the RQD has two electrons and the charging effect due to the Coulomb interaction should appear. Therefore, the SE initially confined in the RQD occupies a two-electron ground state $E_{\mathrm{G(R2)}}$ which has an additional energy with about charging energy $E_{\mathrm{C}}$, roughly estimated to be 19 meV (Appendix A). Note that the conduction band bottom in Figs.~\ref{f4}(d)-(g) is fixed for simplicity and it actually rises by $E_{\mathrm{C}}$. On the other hand, since the injected hot SE has a high energy, it occupies a two-electron excited state $E_{\mathrm{E(R2)}}$ in the RQD. The maximum $E_{\mathrm{E(R2)}}$ relative to the detection barrier height at the intersection point of the black dashed lines in Fig.~\ref{f4}(a) is roughly estimated to 0.2 eV. Since an electron occupying the two-electron ground state can be ejected to the drain at the point of intersection, the SE is ejected not only from $E_{\mathrm{E(R2)}}$ but also from $E_{\mathrm{G(R2)}}$.

When the hot SE is ejected to the drain at a rate $\Gamma _{\mathrm{E2}}$ (we call this situation E ejection), the remaining electron occupies $E_{\mathrm{G(R1)}}$ and it can not be ejected to the drain [Fig.~\ref{f4}(f)]. On the other hand, when the ground-state SE is ejected to the drain with a rate of $\Gamma _{\mathrm{G2}}$ (we call this situation G ejection), the hot-SE energy is reduced by $E_{\mathrm{C}}$. We denote this energy level as $E_{\mathrm{E(R1)}}$. Since Figs.~\ref{f4}(b) and (c) show current saturation, we expect that this hot SE is ejected to the drain in every cycle. Since $\Delta E-E_{\mathrm{C}}\sim 0.18$ eV, this expectation is reasonable considering the experimental result for the zero-electron case in Fig.~\ref{f3}(c), where ejection in every cycle is achieved with a hot-SE energy of about 0.15 eV. Note that the number of SEs in the RQD is reset in the next RQD-pump cycle and the situation returns to that of Fig.~\ref{f4}(d).

In Fig.~\ref{f4}(b), the current level is $ef$ near $V_{\mathrm{inj}}=-0.5$ V, at which there is only current due to the RQD pump. Since this fact indicates that errors during the capture stage of the RQD pumping is negligible, $\Delta I_{\mathrm{P}}$ only originates from the G ejection. From the above consideration, a current level at the intersection point of the black dashed lines in Fig.~\ref{f4}(a) can be expressed as $I_{\mathrm{P}}/ef = 1+\Gamma_{\mathrm{G2}}/(\Gamma_{\mathrm{E2}}+\Gamma_{\mathrm{G2}})=1+\Delta I_{\mathrm{P}}/ef$. From this equation, $\Gamma _{\mathrm{G2}}/\Gamma _{\mathrm{E2}}\sim 0.06$. This low ratio is probably due to the large energy difference of the two SEs.

Now, let us examine the energy changes in the saturation regime using the estimated conversion factors from the voltage to the barrier height or to the RQD energy (see Appendix A and Tab.~\ref{tab:price}). The voltage window in the saturation regime in Fig.~\ref{f4}(b) and (c) is roughly 50 mV. In the case of the $V_{\mathrm{tilt}}$ line, a 50-mV change in $V_{\mathrm{inj}}$ accompanies a 58-mV change in $V_{\mathrm{det}}$. Therefore, $\Delta E$ changes by $0.058\alpha _{\mathrm{detB}}-0.05\alpha_{\mathrm{injB}}\sim 3$ meV. This change is negligibly small compared with the hot-SE energy of about 0.15 eV. On the other hand, in the case of the $V_{\mathrm{inj}}$ line, a 50-mV change in $V_{\mathrm{inj}}$ leads to a change in $\Delta E$ of $0.05(\alpha _{\mathrm{injB}}-\alpha_{\mathrm{inj\_detB}})\sim 19$ meV and to a change in the barrier height relative to the RQD ground state by $0.05(\alpha _{\mathrm{inj\_RQD}}-\alpha_{\mathrm{inj\_detB}})\sim 3$ meV. The former value is not small but, if we subtract $E_{\mathrm{C}}$ from $\Delta E$, we get about 0.16 eV, which indicates the possibility that the hot SE was ejected to the drain in every cycle. The latter value is much smaller than $E_{\mathrm{C}}$. These considerations support the validity of the above model.

\subsection{Hot-SE injection into a two-electron RQD}
Now, let examine the results of hot-SE injection into the RQD with two electrons. To measure current in this condition, we reduced $A_{\mathrm{S}}$, resulting in a lowering (raising) of the Fermi level in the source during the LQD (RQD) pump operation. In this case, we need a more positive $V_{\mathrm{inj}}$ to capture an SE by the LQD. On the other hand, more SEs can be captured by the RQD without changing $V_{\mathrm{det}}$. In addition, SE ejection over the detection barrier is not sensitive to the Fermi level of the source. These aspects suggest that the region of hot-SE injection overlaps with that of the two-SE capture by the RQD in a pumping current map.

\begin{figure}
\begin{center}
\includegraphics[pagebox=artbox]{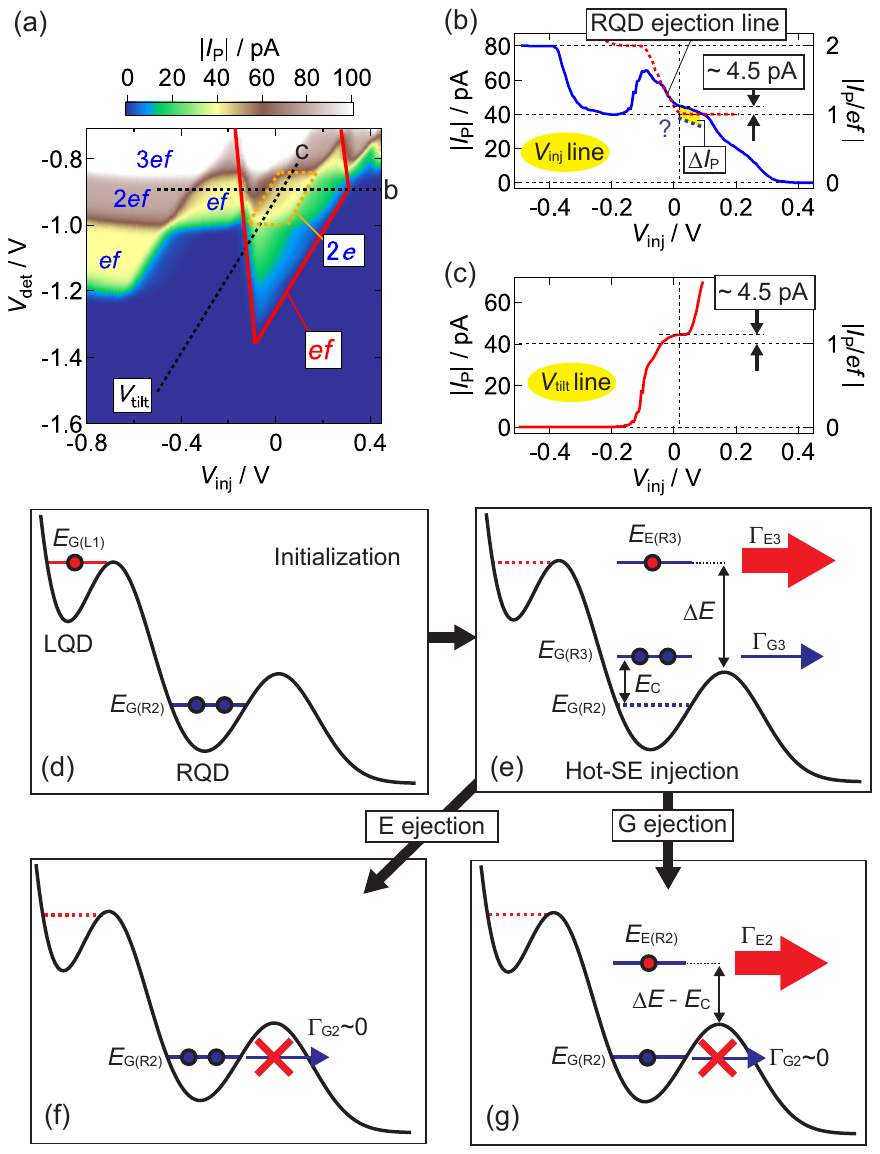}%width=230pt 
\end{center}
\caption{(a) $|I_{\mathrm{P}}|$ as a function of $V_{\mathrm{inj}}$ and $V_{\mathrm{det}}$ with the pulse sequence shown in Fig.~\ref{f1}(c), where $V_{\mathrm{upper}}=1.2$ V, $V_{\mathrm{ent}}=-0.3$ V, $V_{\mathrm{s}}=0$ V, $A_{\mathrm{ent}}=0.3$ V, $A_{\mathrm{inj}}=0.3$ V, and $A_{\mathrm{s}}=0.075$ V. The slope of $V_{\mathrm{tilt}}$ is 1.16. The labels b and c indicate the measured points plotted in (b) and (c), respectively. (b) $|I_{\mathrm{P}}|$ as a function of $V_{\mathrm{inj}}$ with the same voltage condition as (a) except $V_{\mathrm{det}}=-0.895$ V, corresponding to the horizontal black dashed line in (a). (c) $|I_{\mathrm{P}}|$ as a function of $V_{\mathrm{inj}}$ with the same voltage condition as (a) except $V_{\mathrm{det}}=1.16V_{\mathrm{inj}}-0.924$ V, corresponding to the tilted black dashed line in (a). (d)-(g) Potential diagrams for the explanation about the hot-SE injection into the two-electron RQD. The conduction band bottom is fixed for simplicity.}
\label{f5}
 \end{figure}

Figure~\ref{f5}(a) shows such a map as a function of $V_{\mathrm{inj}}$ and $V_{\mathrm{det}}$. The region inside the orange dashed lines corresponds to two electrons in the RQD and the region inside the red lines corresponds to hot-SE ejection from the LQD. Figures~\ref{f5}(b) and (c) plot the current along two black dashed lines in (a). The current level exceeds $ef$ by about 4.5 pA at the intersection point of the two black dashed lines in Fig.~\ref{f5}(a). To indicate where the ejection line of the individual RQD pump is, Fig.~\ref{f5}(b) (red dashed line) plots the RQD pumping characteristics along the red line in Fig.~\ref{f2}(c), with an vertical offset of $ef$ and a horizontal offset determined by matching the slope. Since the direct pumping current from the RQD before the injection of the hot SE is almost zero at the point of intersection, the current of about 4.5 pA is at least related to ejection of the two cold SEs from the RQD as the hot SE propagates through it. The blue dashed line is a non-quantitative guide for the eye showing the current due to the hot SE. $\Delta I_{\mathrm{P}}$, which is implicitly defined as the excess current due to the cold SEs, in the yellow region is larger than 4.5 pA. Note that the hot SE was not ejected in every cycle in this case because its maximum energy was only about 0.13 eV. In addition, there is an excess current of 4.5 pA  along $V_{\mathrm{tilt}}$ but its saturation is not so clear compared with that in Fig.~\ref{f4}(c). This result is reasonable because the ejection probability of the injected hot SE depends on $V_{\mathrm{inj}}$ and $V_{\mathrm{det}}$ in this regime.

Figures~\ref{f5}(d)-(g) show potential diagrams in the case of the two-electron RQD. Here, the conduction band bottom is depicted as being constant for simplicity. The one-electron ground state $E_{\mathrm{G(L1)}}$ in the LQD and two-electron ground state $E_{\mathrm{G(R2)}}$ in the RQD are initially occupied [Fig.~\ref{f5}(d)]. Then, as shown in Fig.~\ref{f5}(e), the injection of a hot-SE into the RQD leads to a charging effect similar to that in Fig.~\ref{f4}(e). Since there are two electrons in the ground state of the three-electron RQD, the rate $\Gamma _{\mathrm{G3}}$ of ejection from the ground state $E _{\mathrm{G3(R3)}}$ is at least two times higher than $\Gamma _{\mathrm{G2}}$. This is consistent with the fact that $\Delta I_{\mathrm{P}}$ ($> 4.5$ pA) is more than twice as large as in the case of the one-electron RQD ($\Delta I_{\mathrm{P}}\sim 2.2$ pA). This consistency also shows the validity of our model.

The remaining processes are essentially the same as those of the one-electron-RQD case. When the hot SE is ejected to the drain at a rate $\Gamma _{\mathrm{E3}}$, two electrons remain trapped in the RQD [Fig.~\ref{f5}(f)]. When one of the ground-state electrons is ejected to the drain at a rate $\Gamma _{\mathrm{G3}}$, only remaining hot electron can be ejected to the drain [Fig.~\ref{f5}(g)]. Note that $\Delta E -E_{\mathrm{C}}\sim 0.11$ eV is insufficient for the hot SE to be ejected in every cycle. Similar to Fig.~\ref{f4}(b), there is a $2ef$ plateau on the left side of Fig.~\ref{f5}(b), indicating negligible errors of the two-SE capture by the RQD.

\subsection{Additional discussions}
Above, we explained the experimental results by using a simple single-particle model, where we assumed that the initially captured SE in the RQD occupies the ground state. However, there is a possibility of transition of the SE from the ground to excited states in the RQD during the hot-SE propagation over the RQD. This process should accompany relaxation of the injected hot SE. If it occurs, the energy distribution of the initially confined SE becomes broad and the SE might be ejected to drain after the E ejection [Figs.~\ref{f4}(f) and \ref{f5}(f)]. Since the match of the RQD ejection lines in Fig.~\ref{f5}(b) is rather good, we expect that the transition is not so large in this voltage regime. This is possibly because the interaction time is too short for the transition. We should note that, although the energy distribution of the injected SE can usually be used to detect the broadening, the distribution of the injected-SE energies is broad [Fig.~\ref{f3}(c)] and it is impossible to distinguish the intrinsic broadening from the broadening due to the relaxation process. Therefore, we can not exclude the possibility of a transition in the voltage regime where the direct RQD pumping current is suppressed. The previous many-electron collision experiment \cite{OnoHot} showed that such a transition frequently occurs and leads to the current amplification. In that case, the number of electrons interacting with an injected hot electron is on the order of one hundred, which might be a reason for the difference between our results and those of the previous work. The transition should be further investigated in future experiments using an RQD with a different length and an injected SE with a sharper distribution.

Regarding flying qubit applications, the charging effect is necessary to perform the two-qubit gate operation \cite{Fly2qb} but transitions between energy levels must not occur. In that sense, the small transition between the energy levels would be good characteristics. In addition, it would be valuable to roughly estimate an expected time evolution of the phase of the two SE system \cite{SAWcol}, which is $E_{\mathrm{C}}\tau/2\hbar \sim 18\pi$, where $\tau\sim 4$ ps is the propagation time of the injected hot SE in the RQD (we roughly use the period of coherent oscillations previously extracted from the data of our SE pump \cite{GYnnano}) and $\hbar$ is the reduced Planck constant. Note that $E_{\mathrm{C}}/2$ is a rough estimate of the electrostatic energy. Since $18\pi$ is too large, this rough estimate suggests that the two SEs should be farther apart. In future experiments, the design of the device should be improved for performing appropriate qubit operations.

\section{Conclusion}
We investigated the injection of hot SEs into zero-, one-, and two-electron QDs by using tunable-barrier series-coupled SE pumps in silicon. In the zero-electron case, the hot SE was ejected in every cycle with a sufficiently high energy (> 0.15 eV) with respect to the detection barrier. The hot-SE energy distribution had a two-step feature possibly due to phonon emission. In the one-electron case, we observed a current in excess of $ef$ when the hot SE is injected. This excess current is attributed to additional ejection of the cold SE trapped in the RQD due to the Coulomb interaction (charging effect). In the two-electron case, the excess current more than doubled, which can be explained by an increase in the ejection rate of the ground-state SEs. These results indicate the existence of a strong unscreened Coulomb interaction and reveal a possibility of building a two-qubit gate with flying electrons in silicon. 

\begin{acknowledgements}
We thank T. Shimizu and K. Nishiguchi for fruitful discussions. This work was partly supported by JSPS KAKENHI Grant Number JP18H05258.
\end{acknowledgements}

\appendix \section{Estimation of conversion factors}
\begin{figure}
\begin{center}
\includegraphics[pagebox=artbox]{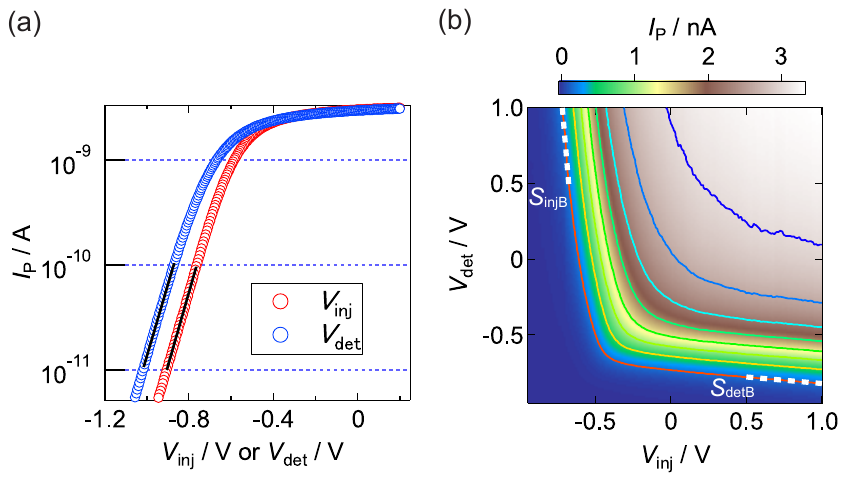}%width=230pt 
\end{center}
\caption{(a) DC current as a function of $V_{\mathrm{inj}}$ ($V_{\mathrm{det}}$) at room temperature (295 K), where $V_{\mathrm{UG}}=1.2$ V, $V_{\mathrm{inj}}=1$ V ($V_{\mathrm{det}}=1$ V), $V_{\mathrm{ent}}=1$ V, and $V_{\mathrm{s}}=1$ mV. The black lines are linear fits to $\mathrm{ln}(I_{\mathrm{P}})$. (b) DC current as a function of $V_{\mathrm{inj}}$ and $V_{\mathrm{det}}$ at room temperature (295 K), where $V_{\mathrm{UG}}=1.2$ V, $V_{\mathrm{ent}}=1$ V, and $V_{\mathrm{s}}=1$ mV. The current level of the contour lines are from 200 pA (red line) to 3 nA (blue line) in steps of 400 pA. The white lines are linear fits to the red contour line. The slopes are $S_{\mathrm{injB}}=-12$ and $S_{\mathrm{injB}}=-0.084$ for the injection and detection barrier, respectively.}
\label{fA1}
 \end{figure}

To estimate the energy of the electrons, we must determine factors for converting $V_{\mathrm{inj}}$ ($V_{\mathrm{det}}$) into an energy value. First, let us focus on the conversion from the applied voltage into the barrier height ($\alpha _{\mathrm{injB}}$ and $\alpha _{\mathrm{detB}}$). Figure~\ref{fA1}(a) shows the current flowing through the injection (detection) barrier as a function of $V_{\mathrm{inj}}$ ($V_{\mathrm{det}}$) at room temperature. From the fits to the data (black lines), which are $\mathrm{exp}(\alpha _{\mathrm{injB}}V_{\mathrm{inj}}/kT+A)$ and $\mathrm{exp}(\alpha _{\mathrm{detB}}V_{\mathrm{det}}/kT+A)$, where $k$ is the Boltzmann constant, $T$ is temperature, and $A$ is a constant, we obtain $\alpha _{\mathrm{injB}}=0.41$ eV/V and $\alpha _{\mathrm{detB}}=0.40$ eV/V.

Next, let us consider the cross coupling. $V_{\mathrm{inj}}$ ($V_{\mathrm{det}}$) slightly changes the detection (injection) barrier. Figure~\ref{fA1}(b) shows $I_{\mathrm{P}}$ as a function of $V_{\mathrm{inj}}$ and $V_{\mathrm{det}}$ at room temperature. The slopes $S_{\mathrm{injB}}$ and $S_{\mathrm{detB}}$ of the linear fits to the red line indicating a current level of 200 pA can be used to determine the cross coupling from $V_{\mathrm{det}}$ to the injection barrier $\alpha _{\mathrm{det\_injB}}=\alpha _{\mathrm{injB}}/S_{\mathrm{injB}}=0.034$ eV/V and from $V_{\mathrm{inj}}$ to the detection barrier $\alpha _{\mathrm{inj\_detB}}=\alpha _{\mathrm{detB}}S_{\mathrm{detB}}=0.034$ eV/V.

\begin{figure}
\begin{center}
\includegraphics[pagebox=artbox]{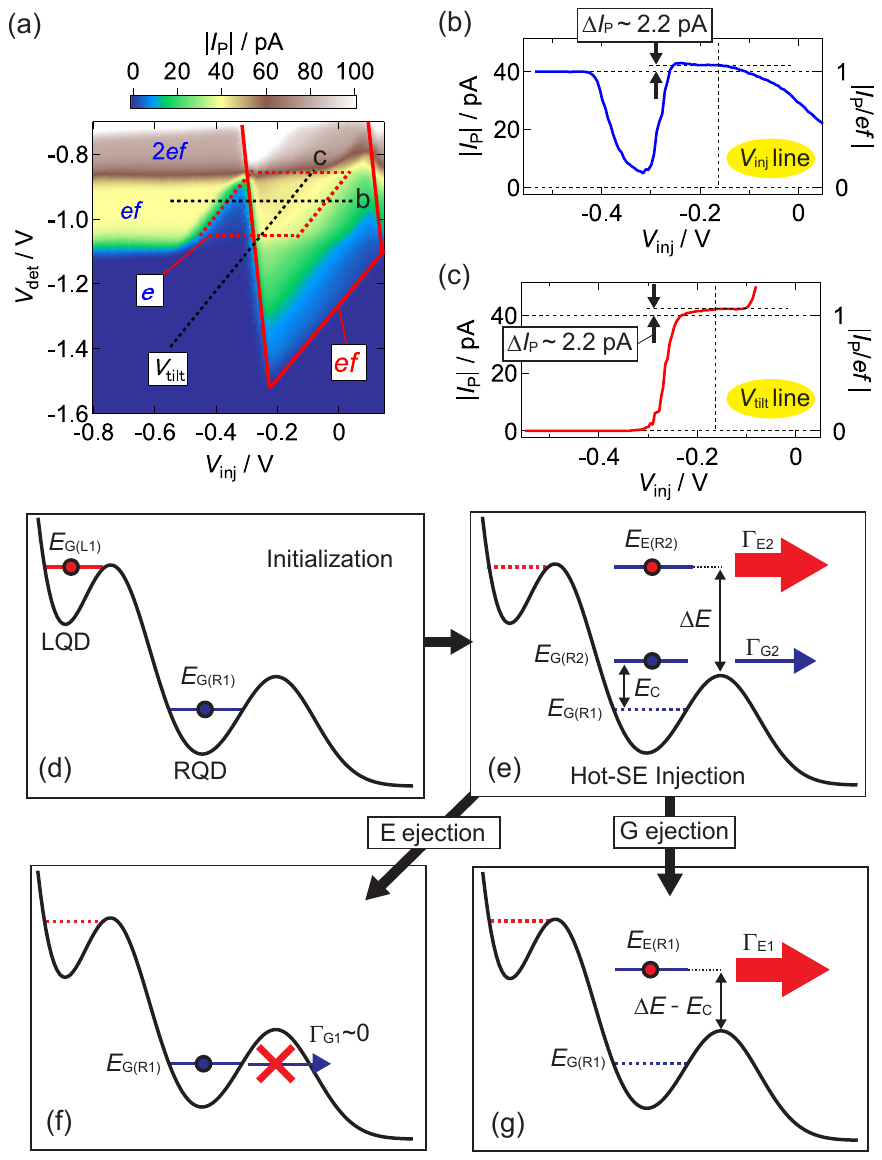}%width=230pt 
\end{center}
\caption{(a) $|I_{\mathrm{P}}|$ as a function of $V_{\mathrm{inj}}$ and $V_{\mathrm{det}}$ with the pulse sequence shown in Fig.~\ref{f1}(c), where $V_{\mathrm{upper}}=1.2$ V, $V_{\mathrm{ent}}=-0.5$ V, $V_{\mathrm{s}}=0$ V, $A_{\mathrm{ent}}=0.3$ V, $A_{\mathrm{inj}}=0.3$ V, and $A_{\mathrm{s}}=0.3$ V. The slope of $V_{\mathrm{tilt}}$ is 1.16. The labels b and c indicate the measured points plotted in (b) and (c), respectively. (b) $|I_{\mathrm{P}}|$ as a function of $V_{\mathrm{inj}}$ with the same voltage condition as (a) except $V_{\mathrm{det}}=-0.945$ V, corresponding to the horizontal black dashed line in (a). (c) $|I_{\mathrm{P}}|$ as a function of $V_{\mathrm{inj}}$ with the same voltage condition as (a) except $V_{\mathrm{det}}=1.16V_{\mathrm{inj}}-0.757$ V, corresponding to the tilted black dashed line in (a). (d)-(g) Potential diagrams for the explanation about the hot-SE injection into the one-electron RQD. The conduction band bottom is fixed for simplicity.}
\label{f4}
 \end{figure}

\begin{figure}
\begin{center}
\includegraphics[pagebox=artbox]{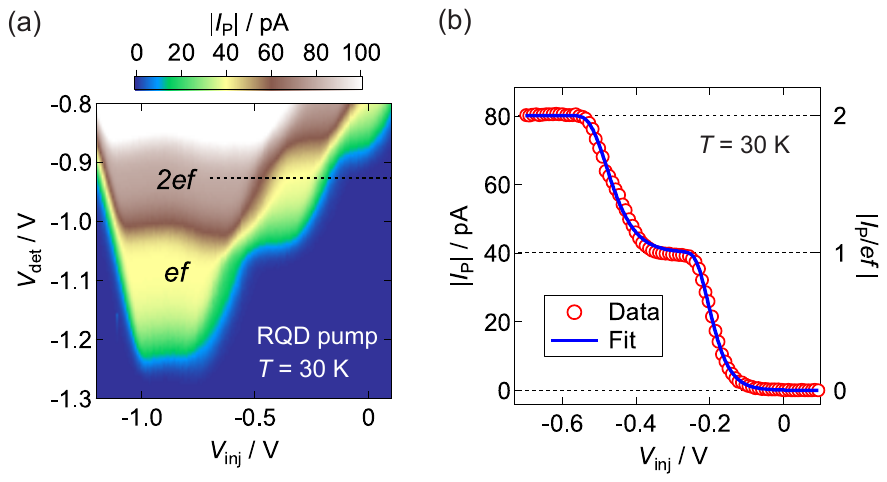}%width=230pt 
\end{center}
\caption{(a) $|I_{\mathrm{P}}|$ as a function of $V_{\mathrm{inj}}$ and $V_{\mathrm{det}}$ at $f=250$ MHz and $T=30$ K, where $V_{\mathrm{upper}}=1.2$ V, $V_{\mathrm{ent}}=1$ V,  and $V_{\mathrm{s}}=0$ V. The high-frequency signal is only applied to the center gate with an amplitude $A_{\mathrm{inj}}=0.3$ V. (b) $|I_{\mathrm{P}}|$ (red circles) as a function of $V_{\mathrm{inj}}$ at $f=250$ MHz and $T=30$ K, where $V_{\mathrm{upper}}=1.2$ V, $V_{\mathrm{det}}=-0.925$ V, $V_{\mathrm{ent}}=1$ V, $V_{\mathrm{s}}=0$ V, and $A_{\mathrm{ent}}=0.3$ V, corresponding to the black dashed line in (a). The blue curve is a fit to the data using Eq.~\ref{eq1}. The fit yields  $\alpha_{\mathrm{inj}}^{1}=0.079$ eV/V, $\alpha_{\mathrm{inj}}^{2}=0.058$ eV/V, $V_{1}=-0.20$ V, and $V_{2}=-0.48$ V.}
\label{fA2}
 \end{figure}

Now, let us estimate the charging energy $E_{\mathrm{C}}$ of the RQD when an SE captured by the RQD is ejected to the drain. Figure~\ref{fA2}(a) shows $|I_{\mathrm{P}}|$ as a function of $V_{\mathrm{inj}}$ and $V_{\mathrm{det}}$ when only the RQD pump is operated at 30 K. Current steps related to the ejection stage appear at the black dashed line in Fig.~\ref{fA2}(a) as shown in Fig.~\ref{fA2}(b). At this high temperature, the ejection is dominated by a thermal hopping and an equation for the current steps can be formulated as follows \cite{GYnnano}:
\begin{equation}
I=2
-\sum_{n=1}^{2}\mathrm{exp}\left [ -\mathrm{exp}\left \{-\frac{\alpha_{n}^{\mathrm{inj}}\left (V_{\mathrm{inj}}-V_{n} \right)}{kT} \right \} \right ],
\label{eq1}
\end{equation}
where $V_{n}$ is the threshold voltage of the $n$th plateau, $\alpha_{n}^{\mathrm{inj}}$ is equal to $\alpha _{\mathrm{inj\_RQD}}-\alpha_{\mathrm{inj\_detB}}$ for the $n$th plateau, and $\alpha _{\mathrm{inj\_RQD}}$ is the conversion factor from $V_{\mathrm{inj}}$ to the RQD energy. A fit to the data using this equation is shown as the blue curve in Fig.~\ref{fA2}(b). The fit gives $\alpha^{\mathrm{inj}}_{1} \neq \alpha^{\mathrm{inj}}_{2}$, probably due to potential fluctuation. As a crude estimate, we computed the average of the two: $\bar{\alpha}_{\mathrm{inj}}=(\alpha^{\mathrm{inj}}_{1}+\alpha^{\mathrm{inj}}_{2})/2=0.069$ eV/V, which leads to $E_{\mathrm{C}}=\bar{\alpha}_{\mathrm{inj}}(V_{1}-V_{2})=19$ meV. This value is consistent with those for previous devices \cite{GYnnano,Nathan_cur, myapl1}. In addition, $\alpha _{\mathrm{inj\_RQD}}=\bar{\alpha}_{\mathrm{inj}}+\alpha_{\mathrm{inj\_detB}}=0.10$ eV/V. Furthermore, from the slope $S=1.16$ of $V_{\mathrm{tilt}}$ [see Fig.~\ref{f4}(a)], we determined the conversion factor from $V_{\mathrm{det}}$ to the RQD ($\alpha _{\mathrm{det\_RQD}}$), i.e., $\alpha _{\mathrm{det\_RQD}}=\alpha _{\mathrm{detB}}-\bar{\alpha}_{\mathrm{inj}}/S=0.34$ eV/V \cite{GYnnano}. All of the conversion factors are summarized in Table~\ref{tab:price}. They indicate that the gate-to-barrier couplings are symmetric for the two gates and the position of the RQD is close to the detection barrier, which is consistent with the previous estimation for a different device \cite{GYnnano}.

\begin{table}[htb]
  \begin{center}
    \caption{Summary of the conversion factors and charging energy}
    \begin{tabular}{||c|c||c|c||} \hline
      $\alpha _{\mathrm{injB}}$&   0.41 eV/V   & $\alpha _{\mathrm{inj\_RQD}}$ & 0.10 eV/V\\
      $\alpha _{\mathrm{detB}}$&  0.40 eV/V  & $\alpha _{\mathrm{det\_RQD}}$ & 0.34 eV/V\\
      $\alpha _{\mathrm{inj\_detB}}$ & 0.034 eV/V  &$E_{\mathrm{C}}$  & 19 meV\\
      $\alpha _{\mathrm{det\_injB}}$ & 0.034 eV/V& -  & -\\ \hline
    \end{tabular}
    \label{tab:price}
  \end{center}
\end{table}

\clearpage

\end{document}